\documentclass[fleqn,usenatbib]{mnras}

\usepackage{newtxtext,newtxmath}

\usepackage[T1]{fontenc}

\DeclareRobustCommand{\VAN}[3]{#2}
\let\VANthebibliography\thebibliography
\def\thebibliography{\DeclareRobustCommand{\VAN}[3]{##3}\VANthebibliography}

\usepackage{graphicx}	
\usepackage{amsmath}	
\usepackage{caption}
\usepackage{subcaption}

\usepackage[normalem]{ulem}
\usepackage{xcolor}
\usepackage{orcidlink}
\usepackage{threeparttable}

\title[(W)acko! A SN progenitor via {\sl JWST}]{Identifying the SN 2022acko progenitor with {\sl JWST}}

\author[S. D. Van Dyk et al.]{
Schuyler D.~Van Dyk,\orcidlink{0000-0001-9038-9950}$^{1}$\thanks{E-mail: vandyk@ipac.caltech.edu (SVD)}
K.~Azalee Bostroem,\orcidlink{0000-0002-4924-444X}$^{2}$\thanks{LSSTC Catalyst Fellow}
WeiKang Zheng,\orcidlink{0000-0002-2636-6508}$^{3}$
Thomas G.~Brink,\orcidlink{0000-0001-5955-2502}$^{3}$
Ori D.~Fox,\orcidlink{0000-0003-2238-1572}$^{4}$
\newauthor
Jennifer E.~Andrews,\orcidlink{0000-0003-0123-0062}$^{5}$
Alexei V.~Filippenko,\orcidlink{0000-0003-3460-0103}$^{3}$
Yize Dong,\orcidlink{0000-0002-7937-6371}$^{6}$
Emily Hoang,\orcidlink{0000-0003-2744-4755}$^{6}$
Griffin Hosseinzadeh,\orcidlink{0000-0002-0832-2974}$^{2}$
\newauthor
Daryl Janzen,\orcidlink{0000-0003-0549-3281}$^{7}$
Jacob E.~Jencson,\orcidlink{0000-0001-5754-4007}$^{8}$
Michael J.~Lundquist,\orcidlink{0000-0001-9589-3793}$^{9}$
Nicolas Meza,\orcidlink{0000-0002-7015-3446}$^{6}$
Dan Milisavljevic,\orcidlink{0000-0002-0763-3885}$^{10,11}$
\newauthor
Jeniveve Pearson,\orcidlink{0000-0002-0744-0047}$^{2}$
David J.~Sand,\orcidlink{0000-0003-4102-380X}$^{2}$
Manisha Shrestha,\orcidlink{0000-0002-4022-1874}$^{2}$
Stefano Valenti,\orcidlink{0000-0001-8818-0795}$^{6}$
D. Andrew Howell\orcidlink{0000-0003-4253-656X}$^{12,13}$
\\
$^{1}$Caltech/IPAC, Mailcode 100-22, Pasadena, CA 91125, USA\\
$^{2}$Steward Observatory, University of Arizona, 933 North Cherry Avenue, Tucson, AZ 85721, USA\\
$^{3}$Department of Astronomy, University of California, Berkeley, CA 94720-3411, USA\\
$^{4}$Space Telescope Science Institute, 3700 San Martin Drive, Baltimore, MD 21218, USA\\
$^{5}$Gemini Observatory/NSF's NOIRLab, 670 N. A'ohoku Place, Hilo, HI 96720, USA\\
$^{6}$Department of Physics and Astronomy, University of California, Davis, 1 Shields Avenue, Davis, CA 95616-5270, USA\\
$^{7}$Department of Physics \& Engineering Physics, University of Saskatchewan, 116 Science Pl, Saskatoon, SK S7N 5E2, Canada\\
$^{8}$Department of Physics and Astronomy, The Johns Hopkins University, Baltimore, MD 21218, USA\\
$^{9}$W.~M.~Keck Observatory, 65-1120 Mamalahoa Highway, Kamuela, HI 96743, USA\\
$^{10}$Department of Physics and Astronomy, Purdue University, 525 Northwestern Avenue, West Lafayette, IN 47907, USA\\
$^{11}$Integrative Data Science Initiative, Purdue University, West Lafayette, IN 47907, USA\\
$^{12}$Las Cumbres Observatory, 6740 Cortona Dr. Suite 102, Goleta, CA 93117, USA\\
$^{13}$Physics Department, University of California, Santa Barbara, Santa Barbara, CA 93111, USA\\
}

\date{Accepted 2023 June 29. Received 2023 June 29; in original form 2023 February 01}

\pubyear{2023}

\begin{document}
\label{firstpage}
\pagerange{\pageref{firstpage}--\pageref{lastpage}}
\maketitle

\begin{abstract}
We report on analysis using the {\sl James Webb Space Telescope\/} ({\sl JWST}) to identify a candidate progenitor star of the Type II-plateau supernova SN~2022acko in the nearby, barred spiral galaxy NGC 1300. To our knowledge, our discovery represents the first time {\sl JWST\/} has been used to localize a progenitor system in pre-explosion archival {\sl Hubble Space Telescope\/} ({\sl HST}) images. We astrometrically registered a {\it JWST\/} NIRCam image from 2023 January, in which the SN was serendipitously captured, to pre-SN {\sl HST\/} F160W and F814W images from 2017 and 2004, respectively. An object corresponding precisely to the SN position has been isolated with reasonable confidence. That object has a spectral energy distribution and overall luminosity consistent with a single-star model having an initial mass possibly somewhat less than the canonical 8~M$_{\odot}$ theoretical threshold for core collapse (although masses as high as 9~M$_{\odot}$ for the star are also possible); however, the star's SED and luminosity are inconsistent with that of a super-asymptotic giant branch star which might be a forerunner of an electron-capture SN. The properties of the progenitor alone imply that SN~2022acko is a relatively normal SN~II-P, albeit most likely a low-luminosity one. The progenitor candidate should be confirmed with follow-up {\sl HST\/} imaging at late times, when the SN has sufficiently faded. This potential use of {\sl JWST\/} opens a new era of identifying SN progenitor candidates at high spatial resolution.
\end{abstract}




\section{Introduction}

Supernovae (SNe) are the catastrophic endpoints of the lives of some stars. It is widely believed that the total disruption of a white dwarf (WD), as a consequence of critical mass accretion from a dwarf or giant star (or a merger with another WD), leads to thermonuclear explosion as  normal Type Ia SNe, which have proven to be highly valuable cosmological probes owing to their high, standardizable luminosity. Stars with initial masses $M_{\rm ini} \gtrsim 8$~M$_{\odot}$ reach a point in their evolution at which nuclear burning can no longer support the inner core, and the collapse of the core to neutron degeneracy rapidly leads to the explosive removal of the outer stellar envelope --- a core-collapse SN (CCSN; e.g., \citealt{Woosley1986}). CCSNe account for $\sim 76$\% of all SNe locally \citep{Li2011,Graur2017}. CCSNe are not homogeneous in their properties: both spectroscopically and photometrically they roughly divide further into the H-rich Type II and the H-poor Type Ib and Ic, with Type IIb constituting an intermediate grouping \citep[see, e.g.][for reviews of SN classification]{Filippenko1997,GalYam2017}. Type II SNe have been classically separated into the Type II-plateau (II-P) and II-linear (II-L), based on their light-curve properties, although such distinctions in the modern view are debatable \citep{Anderson2014,Valenti2016}. Further subdivisions have emerged among the SNe~II, Ib, and Ic with respect to the presence of relatively narrow spectral emission features (the so-called ``n'' subtypes), interpreted as being indicative of the presence of dense circumstellar material.

Of utmost importance in understanding SNe in general, and CCSNe specifically, is mapping the various SN subtypes to classes of progenitor star. A number of indirect methods have been employed, but the most direct approach is to identify the actual star that has exploded. The lion's share of such detections have been made in pre-explosion {\sl Hubble Space Telescope\/} ({\sl HST}) archival images. As a community, we have successfully demonstrated in more than 20 cases that SNe~II-P, in particular, arise from massive stars in the red supergiant (RSG) phase, as expected theoretically \citep[][]{Smartt2009,Smartt2015,VanDyk2017}. The RSGs are likely in a limited range of $M_{\rm ini}$, the low end at the core-collapse limit and $\lesssim 17$~M$_{\odot}$ at the high end (although see \citealt{Davies2020}). SNe~IIn appear to arise potentially from a higher-mass stellar group, with remarkable outbursts observed before explosion, ostensibly as luminous blue variable stars \citep[e.g.,][]{GalYam2009,Smith2022,Brennan2022,Jencson2022}. For SNe~Ib, a hot He-star progenitor was identified for one event (iPTF13bvn; e.g., \citealt{Eldridge2016}), whereas a cool, extended star was identified for another (SN~2019yvr; \citealt{Kilpatrick2021}). A potentially hot, luminous candidate progenitor has been isolated for one SN~Ic (SN~2017ein; \citealt{VanDyk2018,Kilpatrick2018,Xiang2019}), although this remains to be confirmed.

The most prevalent CCSNe, SNe~II-P ($\sim 48$\% locally; \citealt{Smith2011}), are themselves also quite heterogeneous, most notably with a range in both peak and overall luminosities, as well as plateau duration \citep[e.g.,][]{Anderson2014,Valenti2016,deJaeger2019}. At the low-luminosity end, the RSG progenitors have all been found to be at the lower end of the $M_{\rm ini}$ range, such as SN~2005cs ($9^{+3}_{-2}$ M$_{\odot}$, \citealt{Maund2005}; $10 \pm 3$ M$_{\odot}$, \citealt{Li2006}), SN~2008bk (8--8.5 M$_{\odot}$, \citealt{VanDyk2012}; 8--10 M$_{\odot}$, \citealt{ONeill2021}), and SN~2018aoq ($\sim 10$ M$_{\odot}$, \citealt{ONeill2019}). For a nominal local Universe initial-mass function, we would expect lower-mass progenitors and, thus, low-luminosity SNe~II to predominate in number. What makes this of critical interest is that these low masses are very near the expected boundaries between WD formation and core collapse, as well as the intermediate scenario of electron-capture (EC) explosions of super-asymptotic giant branch (AGB) stars with O-Ne-Mg cores \citep[e.g.,][]{Miyaji1980,Nomoto1984,Doherty2017}. The recent case of SN~2018zd has brought this potential fine line to the fore: \citet{Hiramatsu2021} claimed, based on the SN's properties, this SN was the first {\it bona fide\/} example of an ECSN, whereas \citet{Zhang2020} argued that SN~2018zd had properties more consistent with those of normal SNe~II-P. Both studies characterised the identified progenitor candidate as potentially being a super-AGB (SAGB) star (\citealt{VanDyk2023} have since confirmed the candidate as the progenitor). Nevertheless, probing where these boundaries exist and what factors govern them is a key area of exploration in our understanding of stellar astrophysics and evolution. Garnering further examples of low-luminosity SNe~II-P and their progenitors is therefore important and welcome.

Here we present the identification of the progenitor of the SN~II-P 2022acko. The precise locations of previous SNe in archival images have been established either using {\sl HST\/} or adaptive-optics ground-based observations of the SN itself. In this case, to our knowledge for {\em the first time}, we employ the novel method of pinpointing SN~2022acko in pre-SN {\sl HST\/} data via {\sl JWST\/} images containing the SN. 
SN 2022acko was first discovered on 2022 December 6 (UTC dates are used throughout this paper) at 16.5~mag, $58{\farcs}2$ north and $29{\farcs}6$ west of the nucleus of NGC~1300 \citep{Lundquist2022}, by the Distance Less Than 40 Mpc (DLT40) survey \citep{Tartaglia2018}. It was classified the following day as a young SN~II-P \citep{Li2022}. The SN is also known as DLT22v, ATLAS22bnms, ZTF22abyivoq, PS22mpv, and Gaia23aap. \citet{Bostroem2023} provided a detailed description of the SN at early times, including analysis of {\sl HST\/} ultraviolet spectra; they find that it is most consistent with a low-luminosity SN~II-P. We also present additional limited optical, ground-based follow-up observations in this paper which further corroborate that characterisation. SN 2022acko is, as far as we know, the first historical SN to be discovered in this nearby, nearly face-on barred spiral galaxy. 

\section{Observations and reductions}

\subsection{{\sl HST\/} data}

The site of SN 2022acko was serendipitously covered in various {\sl HST\/} observations of the host galaxy by several different programs. We obtained all of these publicly available data from the Mikulski Archive for Space Telescopes (MAST). The first set of data was obtained with the now-decommissioned Wide-Field and Planetary Camera 2 (WFPC2) on 2001 January 6 (21.9 yr before explosion) in F606W by program GO-8597 (PI M.~Regan). The second set was obtained with the Advanced Camera for Surveys in the Wide Field Channel (ACS/WFC) on 2004 September 21 (18.2 yr) in F435W, F555W, F658N, and F814W by GO/DD-10342 (PI K.~Noll; these were ``Hubble Heritage'' observations); note that only the ``NGC1300-POS1'' field contains the relevant half of the galaxy with the SN site. Next, the host galaxy was observed with the Wide-Field Camera 3 IR channel (WFC3/IR) on 2017 October 27 (5.1 yr) in F160W by GO-15133 (PI P.~Erwin). Observations were also obtained with the WFC3 UVIS channel (WFC3/UVIS) in F475W and F814W; however, these were in subarray mode, centered on the nucleus, so they missed the SN site. Finally, WFC3/UVIS observations were conducted, in full-array mode, on 2020 January 4 (2.9 yr) in F275W and F336W by GO-15654 (PI J.~Lee; PHANGS-HST, \citealt{Lee2022}). We show portions of the image mosaics in F814W and F160W, containing the SN site, in Figure~\ref{fig:triptych}.

\subsection{{\sl JWST\/} data}

The host galaxy was also observed on 2023 January 25, just 52~d after explosion, using {\sl JWST\/} with both the NIRCam and MIRI instruments by program GO-2107 (PI J.~Lee; PHANGS-JWST, \citealt{Lee2023}). The bands used for NIRCam in both the short- and long-wavelength channels were F200W (9620~s), F300M (773~s), F335M (773~s), and F360M (859~s), whereas for MIRI, the bands were F770W, F1000W, F1130W, and F2100W. We obtained these publicly available data as Level-3 mosaics from MAST as soon as they appeared. The focus of this paper is on utilising the NIRCam data for the purpose of potentially identifying the SN progenitor, since those bands match up best in wavelength and at comparable spatial resolution with the available {\sl HST\/} bands, whereas the MIRI images do not meet these stipulations. We will therefore not be analyzing the MIRI data here at all. 

\begin{figure*}
     \centering
     \begin{subfigure}[b]{0.3\textwidth}
         \centering
         \includegraphics[width=\textwidth]{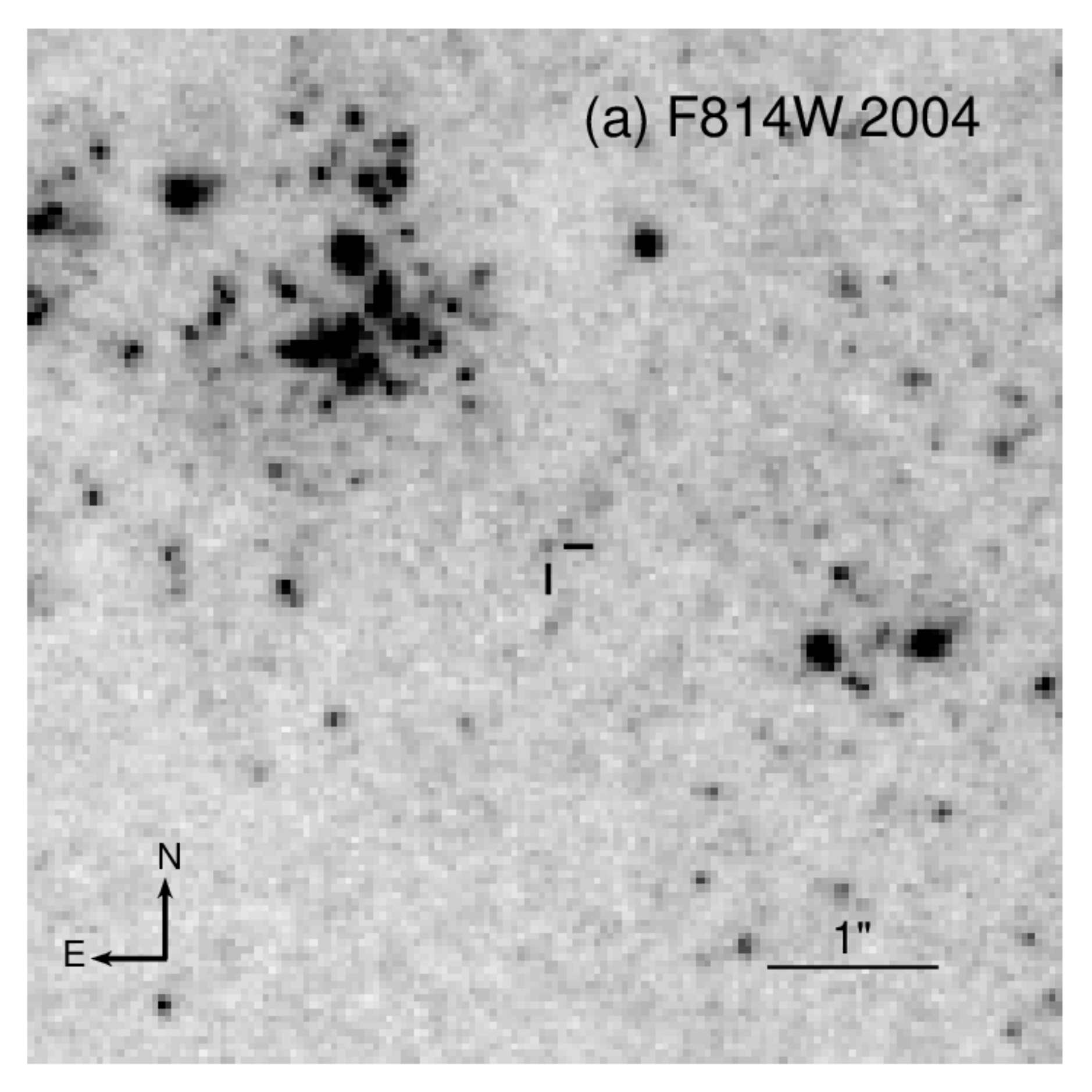}
     \end{subfigure}
     \hfill
     \begin{subfigure}[b]{0.3\textwidth}
         \centering
         \includegraphics[width=\textwidth]{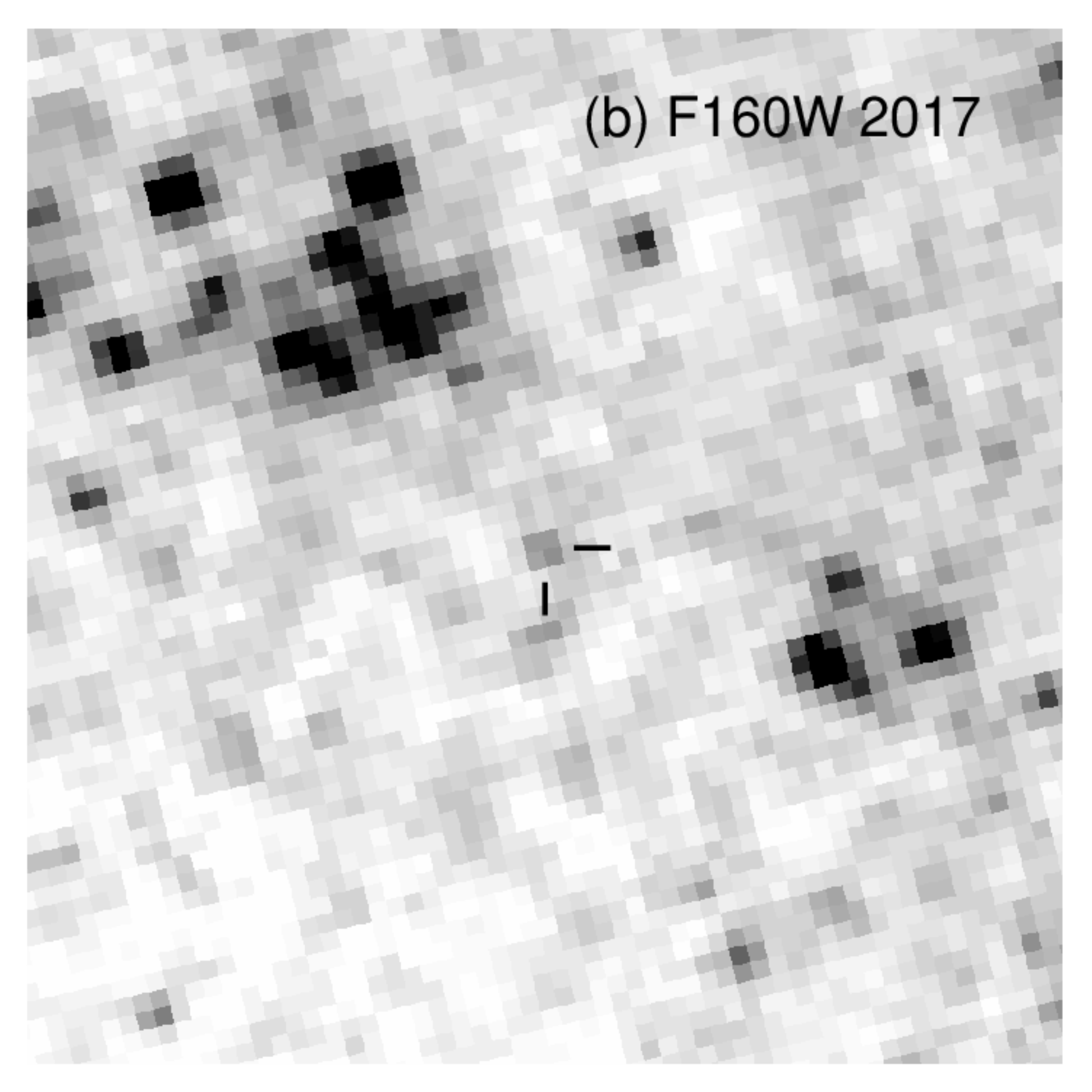}
     \end{subfigure}
     \hfill
     \begin{subfigure}[b]{0.3\textwidth}
         \centering
         \includegraphics[width=\textwidth]{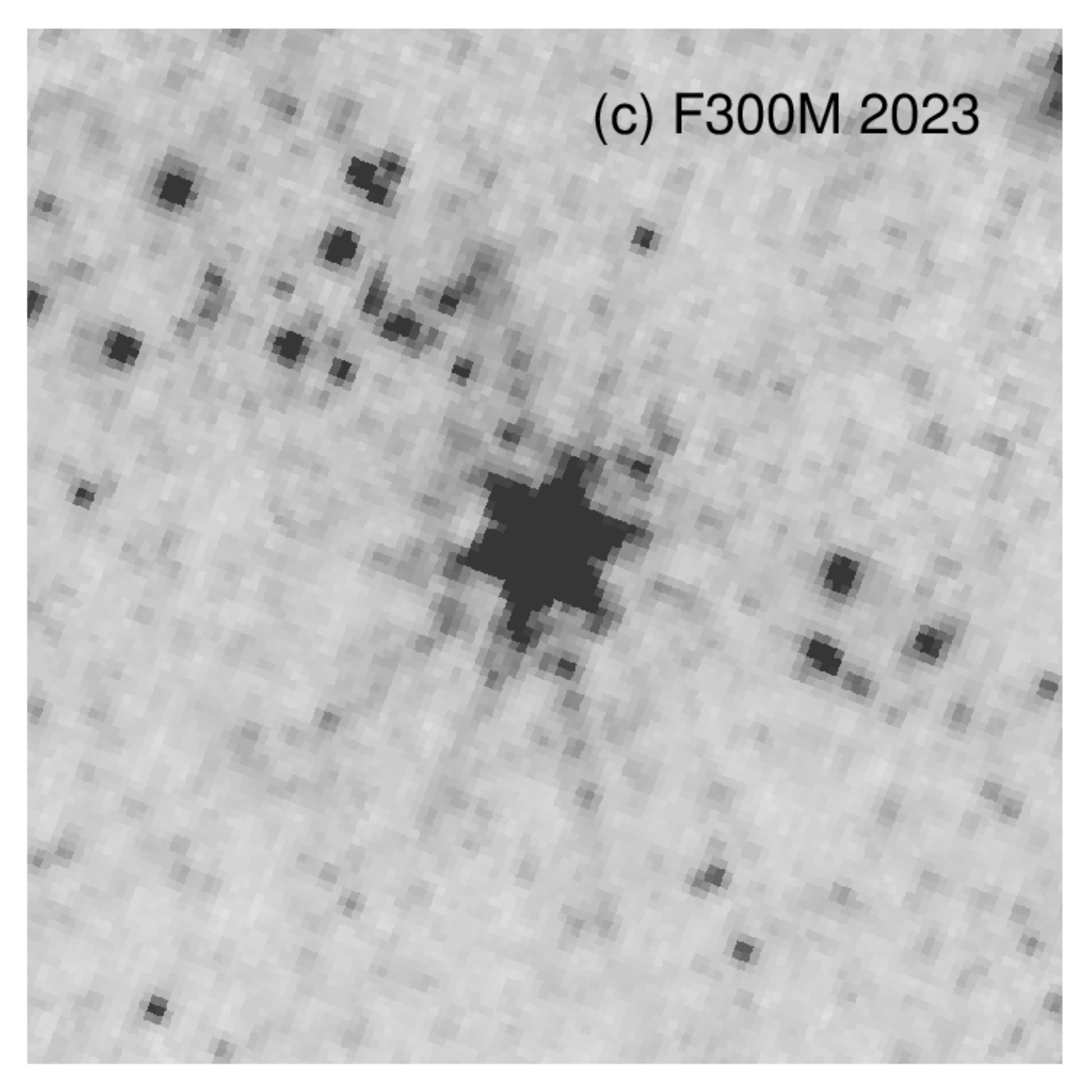}
     \end{subfigure}
        \caption{{\it Left:} A portion of a {\sl HST\/} WFC3/UVIS image mosaic in F814W from 2004 September 21, with the progenitor candidate for SN~2022acko indicated with tick marks. {\it Middle:} same as the left panel, but for a WFC3/IR F160W mosaic from 2017 October 27. {\it Right:} a portion of a {\sl JWST\/} NIRCam mosaic in F300M from 2023 January 25; the SN is $\approx 108$ pixels from the frame edge. All panels are shown at the same scale and orientation. North is up and east is to the left.}
        \label{fig:triptych}
\end{figure*}

The point-spread-function (PSF) profile of the SN in F200W, unfortunately, appears hopelessly saturated, as a result of the SN's brightness and exposure depth of the observations in that band, making it worthless for our purpose. On the contrary, the PSF was of high quality in the remaining medium bands. The first two medium bands are intended to sample various molecular emission features from components in the interstellar medium (water ice, PAHs/CH$_4$), and the third band is intended for continuum emission from brown dwarfs and planets. Nebular emission is apparent in the images in those first two bands; however, we selected F300M for use here, rather than the continuum band (F360M); any ice emission features in the data do not appear to be especially strong, such that most of the observed flux in the image appears to arise from point sources. Furthermore, F300M is the one closest in wavelength to the reddest {\sl HST\/} band (F160W). We show a portion of the image mosaic in F300M, containing the SN, in Figure~\ref{fig:triptych}.

\subsection{{\sl Spitzer} data}

Archival data containing the SN site from the post-cryogenic {\sl Spitzer Space Telescope\/} mission are publicly available from the NASA/IPAC Infrared Science Archive (IRSA), via the {\sl Spitzer\/} Heritage Archive. These data are in the 3.6 and 4.5 $\mu$m bands on 2009 October 4 (13.2 yr) from program 61009 (PI W.~Freedman) and 2009 September 5 (13.3 yr) and October 10 (13.2 yr) from program 61065 (PI K.~Sheth). 

\subsection{Astrometry and photometry}

The {\sl HST\/} WFPC2, ACS/WFC, and WFC3/UVIS data were mosaicked per band using {\tt Astrodrizzle} \citep{STScI2012}. In the case of WFC3/IR, both the pipeline-drizzled ``drz'' mosaic and the individual ``flt'' frames obtained from MAST were utilised. Similarly, we obtained the pipeline-produced {\sl JWST\/} NIRCam ``i2d" mosaic directly from the archive.

In order to perform the necessary relative astrometry and determine which star, if any, could be considered a progenitor candidate, we needed to select a number of relatively isolated stars in common between both the {\sl HST\/} F160W and {\sl JWST\/} F300M mosaics. However, we became aware that inherent astrometric distortion exists for NIRCam, for the long-wavelength channel with both modules, particularly near the frame edges \citep[e.g.,][]{Griggio2023}; even after pipeline correction, coherent distortions can remain across the field-of-view \citep[e.g.,][]{DeFurio2023}. We therefore selected 21 stellar-like fiducial objects in close relative proximity to (within $\approx$23\arcsec\ of) the SN site. We employed the centroiding package within {\tt photutils}\footnote{https://photutils.readthedocs.io/en/stable/index.html} \citep{Bradley_2022}, {\tt photutils.centroids}, on these 21 objects in each image dataset. From the results of each of the four centroiding methods ({\tt centroid{\textunderscore}com}, fitting an object's ``center of mass;'' {\tt centroid{\textunderscore}1dg}, fitting one-dimensional (1D) Gaussians to $X$ and $Y$; {\tt centroid{\textunderscore}2dg}, fitting a 2D Gaussian to the PSF; and, {\tt centroid{\textunderscore}quadratic}, fitting a 2D quadratic polynomial), we can establish a set of average centroids for the fiducials on the NIRCam mosaic, with a 1$\sigma$ root-mean-square (rms) uncertainty of 0.20~pixel (12.6~mas, at $0{\farcs}063$~pixel$^{-1}$), and for the WFC3/IR mosaic, with an rms uncertainty of 0.37~pixel ({$\sim$}48.1~mas, at {$\sim$}$0{\farcs}13$~pixel$^{-1}$). Using the task {\it geomap\/} within {\tt PyRAF} with second-order fitting, from the merged fiducial list we were able to align the two mosaics with an rms uncertainty in the mean astrometry of 0.21~WFC3/IR pixel ({$\sim$}27.3~mas). As mentioned above, the PSF of the SN itself in the F300W mosaic was very stellar-like, and we could therefore measure its centroid in those {\sl JWST\/} data with rms uncertainties (from averaging the results from the four centroiding methods) of 0.08~pixel (5.0~mas), and transform its position with the task {\it geoxytran\/} to the F160W mosaic. Doing so, we were able to isolate the object, as indicated in Figure~\ref{fig:triptych} (middle panel). The overall rms uncertainty in the transformation, adding all of the above uncertainties in quadrature, is estimated to be {$\sim$}57.2 mas, or {$\sim$}0.44 WFC3/IR pixel.

We obtained the centroid of this object in a similar fashion using {\tt photutils}, with an rms uncertainty of 0.35~WFC3/IR pixel ($\sim$45.5~mas). The transformed SN position onto the F160W mosaic differs from the centroid of the candidate by 0.54~WFC3/IR pixel ($\sim$70.6~mas). However, given the uncertainty in the object's centroid, this difference is within the overall uncertainties in the astrometric transformation (above). We therefore have confidence in assigning this object as the progenitor candidate.

We performed a similar alignment between the F814W and F300M mosaics, based on the same fiducials in common. The average centroids of the fiducials in F814W have an uncertainty of 0.25~WFC3/UVIS pixel (9.8~mas, at $0{\farcs}039$~pixel$^{-1}$). The uncertainty in the {\it geomap\/} solution is 0.28~UVIS pixel (10.9~mas). The SN position in the NIRCam mosaic transformed to the WFC3/UVIS mosaic is nearly coincident with the object at F814W indicated in Figure~\ref{fig:triptych} (left panel). The overall uncertainty in this transformation is 20.0~mas, or 0.51~UVIS pixel. The centroid for that object, with an rms uncertainty of 0.15~UVIS pixel (5.9~mas), is 0.87~pixel (33.9~mas) from the transformed SN position. In this case, the object centroid is $\gtrsim 1\sigma$ from the transformed position.

Despite the larger offset of the SN position from the object detected at F814W, we suspect that the objects in both F160W and F814W are the same --- the progenitor candidate. To attempt to verify this, we transformed between the F160W and F814W mosaics using the same fiducials. In this case, the {\it geomap\/} uncertainty is 0.45~UVIS pixel (17.6~mas). The total uncertainty in the transformation from F160W to F814W is 69.2~mas, or 1.78~UVIS pixels, whereas the difference between the transformed F160W candidate position to F814W is 1.55~UVIS pixels. With the uncertainty in the F814W centroid, the difference is again slightly greater than the $1\sigma$ uncertainty; however, we still consider it probable that these two objects are one and the same.

Before leaving this discussion of the astrometry entirely, note that we consider it likely that the SN image was near or at saturation in the F300W mosaic. The NIRCam observations were obtained 51.6~d post-explosion \citep{Bostroem2023}, when the SN was at $V \approx 16.4$~mag (see Sec.~\ref{sec:SCM}). If SN 2022acko is similar to SN 2005cs \citep{Bostroem2023}, for which $V-K \approx 1.5$~mag at about the same age, then $K \approx 14.9$~mag. Using the {\sl JWST\/} exposure-time calculator at this brightness in F200W ($\sim K$), we encounter warnings of pixels experiencing partial and full saturation. Furthermore, this $K$ brightness is within $\sim$0.5 mag of the saturation limits for F300M\footnote{https://jwst-docs.stsci.edu/jwst-near-infrared-camera/nircam-performance/nircam-bright-source-limits}. This tends to indicate that, although the SN PSF appears well-behaved, the centroiding may have been affected by saturated or nearly-saturated pixels near the PSF core.

Finally, at the assumed distance to the host (see below), we cannot rule out that the object is a blend or a compact star cluster (the 1.84~pixel full width at half-maximum intensity [FWHM] of the PSF at F160W corresponds at that distance to $\sim$22 pc). Nevertheless, we proceed from this point under the assumption that this object is the SN progenitor candidate; of course, it remains to be confirmed that this is indeed the progenitor when the candidate is shown to have vanished \citep[e.g.,][]{Maund2009,Maund2014,Maund2015,VanDyk2023}.

As a by-product of running {\tt Astrodrizzle} on the {\sl HST\/} data, cosmic-ray (CR) hits on the detectors are masked in the data-quality array for each frame. The frames for each of the various {\sl HST\/} bands were then run through PSF-fitting photometry using {\tt Dolphot} \citep{Dolphin2016}, with the mosaic in one band serving as the source detection reference. {\tt Dolphot} input parameters were set at FitSky=3, RAper=8, and InterpPSFlib=1, with further charge-transfer efficiency (CTE) correction set to zero (except for the one WFPC2 dataset). The CR-hit flagging is important, in particular, for the aperture-correction measurements on each frame. The photometric results from {\tt Dolphot} are returned as Vega magnitudes, and we present these in Table~\ref{tab:prog_phot}. 

The star is detected only in F814W and F160W. We note that the F160W detection is at a formal signal-to-noise ratio (S/N) of 15.3, and the F814W detection is at S/N $= 12.1$. The star was not detected in any of the other bands.   Instead, we provide upper brightness limits on detection in these bands. Here we set these limits at the formal S/N $=5$ returned by {\tt Dolphot} (following \citealt{VanDyk2023}, but see the limitations and caveats associated with this assumption). 

\begin{table}
	\centering
	\caption{{\sl HST\/} photometry of the SN 2022acko progenitor candidate.}
	\label{tab:prog_phot}
	\begin{tabular}{lc} 
		\hline
		Band & Mag$^a$ \\
		\hline
		F275W & $>25.2$ \\
        F336W & $>26.0$ \\
        F435W & $>27.8$ \\
        F555W & $>27.0$ \\
        F606W & $>26.0$ \\
        F658N & $>24.7$ \\
        F814W & 25.61(09) \\
        F160W & 22.88(07) \\
		\hline
	\end{tabular}
    \begin{tablenotes}
    \item[a] $^a$Magnitudes are in the Vega system, and uncertainties of hundredths of a mag are in parentheses; upper limits are 5$\sigma$.
    \end{tablenotes}
\end{table}

For the {\sl Spitzer\/} data we combined all of the artefact-corrected basic calibrated data (BCD) in each band and mosaicked these frames using \texttt{MOPEX} \citep{Makovoz2005a}. PSF-fitting photometry was also executed on the individual frames using \texttt{APEX} \citep{Makovoz2005b}. No source was visibly detected at the relatively uncrowded SN site in the resulting mosaics (not shown). From the photometry we considered the faintest detected source nearest the SN site ($\lesssim 30\arcsec$) in each band, in order to place upper limits on detection of the progenitor: these were at flux densities of 13.58 (S/N $=9.1$) and 8.70 (S/N $=9.7$) $\mu$Jy at 3.6 and 4.5 $\mu$m, respectively (aperture and pixel-phase corrected; color correction is negligible). These limits correspond to $>18.3$ mag in both {\sl Spitzer\/} bands\footnote{Based on the IRAC zeropoints, $272.2 \pm 4.1$ and $178.7 \pm 2.6$ Jy, at the nominal channel wavelengths of 3.544 and 4.487 $\mu$m, respectively.}.

\section{The SN 2022\lowercase{acko} progenitor}\label{sec:progenitor}

The next step is to interpret the photometry by building a spectral energy distribution (SED) for the progenitor candidate. This is reasonably straightforward to do, in that we only have two photometric detections along with many upper limits. For lack of a reliable Cepheid-based or tip-of-the-red-giant-branch distance indicator, we initially adopt a distance to the host galaxy of $18.99 \pm 2.85$~Mpc (distance modulus $\mu=31.39 \pm 0.33$~mag) that \citet{Anand2021} estimated from the Numerical Actions Method (NAM; \citealt{Shaya2017}). (Although see Sec.~\ref{sec:SCM} below.) Regarding the extinction to SN~2022acko, from an analysis of the Na~{\sc i}~D line profiles in an early-time moderate-resolution spectrum \citet{Bostroem2023} measured essentially the same equivalent width for a component associated internally to the host as a component associated with the Galactic foreground. The foreground visual extinction is $A_V=0.083$~mag, so we assume here that the total extinction, $A_V({\rm tot})$, is equivalent simply to doubling the foreground value: $A_V({\rm tot}) \approx 0.166$~mag. \citeauthor{Bostroem2023}~estimated a slightly higher value, 0.205~mag, but this difference is not particularly critical.

\begin{figure}
	\includegraphics[width=\columnwidth]{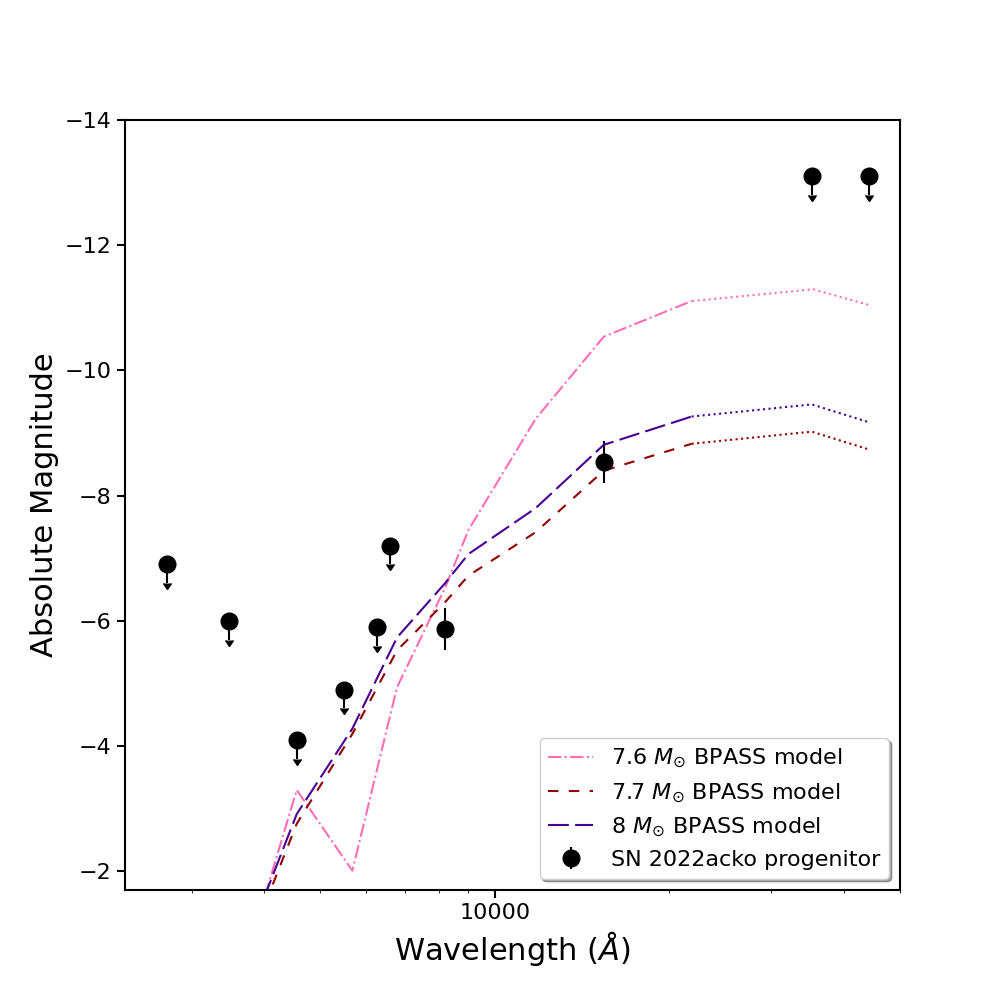}
    \caption{The SN~2022acko progenitor candidate SED, consisting of detections in {\sl HST\/} F814W and F160W, with 5$\sigma$ upper limits to detection in several other {\sl HST\/} bands (see Table~\ref{tab:prog_phot}), as well as $\sim 10\sigma$ upper limits at {\sl Spitzer\/} 3.6 and 4.5 $\mu$m. The observed photometry has been corrected for an assumed distance of $18.99 \pm 2.85$~Mpc \citep{Anand2021} and total visual extinction $A_V=0.166$~mag (assuming $R_V=3.1$ for the reddening correction). Also shown are the endpoint SEDs of theoretical single-star models from BPASS v2.2.2 \citep{Stanway2018} at $M_{\rm ini}=7.6$, 7.7, and 8~M$_{\odot}$. These model SEDs have been extrapolated out to {\sl Spitzer\/} wavelengths using MARCS stellar atmosphere models \citep{Gustafsson2008} at the relevant $T_{\rm eff}$.}
    \label{fig:progenitor_sed}
\end{figure}

Correcting the observed photometry for the distance and extinction, assuming an interstellar reddening law with $R_V=3.1$, results in the SED for the progenitor candidate shown in Figure~\ref{fig:progenitor_sed}. The uncertainties in the F814W and F160W data arise from the uncertainty in the distance modulus, the photometric uncertainties, and the difference in the assumed extinctions between \citet{Bostroem2023} and this study, all added in quadrature; the total 1$\sigma$ uncertainties are overwhelmingly driven by the uncertainty in the distance. 

The final step, then, is to model the observed SED. We first consider recent single-star evolutionary models from BPASS (the Binary Population and Spectral Synthesis code, v2.2.2; \citealt{Stanway2018}). For lack of observational evidence to the contrary, we assume models at solar metallicity, which is nominally consistent with the $\sim$7~kpc distance of the SN site from the host nucleus. We show in Figure~\ref{fig:progenitor_sed} the endpoints of three single-star models at slightly different values of $M_{\rm ini}$: 8, 7.7, and 7.6~M$_{\odot}$. (Note that the termini of the BPASS models are the end of carbon burning.) 

As can be seen in the figure, the upper limits on the SED, including those from {\sl Spitzer}, are not particularly constraining, and the models provide less-than-satisfying overall fits. The 8~M$_{\odot}$ model, an RSG with effective temperature $\log\,T_{\rm eff}/{\rm K} = 3.544$, bolometric luminosity $\log\,(L_{\rm bol}/\mathrm{L}_{\odot})=4.454$, and radius $R=459$~R$_{\odot}$ appears to be too luminous in F814W. The 7.7~M$_{\odot}$ model, with $\log\,T_{\rm eff}/{\rm K}=3.552$, $\log\,(L_{\rm bol}/\mathrm{L}_{\odot})=4.304$, and $R=372$~R$_{\odot}$ is a better comparison, although it is slightly beyond the 1$\sigma$ uncertainty in F814W. For the 7.6 M$_{\odot}$ model, the transition from massive RSG to SAGB has occurred, according to BPASS; this model SED has a markedly different shape and total luminosity (with $\log\,(T_{\rm eff}/{\rm K})=3.475$, $\log\,(L_{\rm bol}/\mathrm{L}_{\odot})=5.047$, and $R=1247$~R$_{\odot}$) than do the RSG models and provides overall a very poor fit to the observations. Note that the model SEDs were extrapolated beyond the bandpasses available from BPASS via synthetic photometry of MARCS model atmospheres \citep{Gustafsson2008} at 3500~K for the RSG models and 3000~K for the SAGB model. Furthermore, following the criteria for CCSN progenitors from \citet{Eldridge2013,Eldridge2017}, specifically of a CO-core mass at explosion with $> 1.38$~M$_{\odot}$, both the 7.6 and 7.7 M$_{\odot}$ models fail to meet all of these criteria, making them less likely in this case; the 8~M$_{\odot}$ model, however, does meet all of these criteria.

We also found that two BPASS binary models had F814W and F160W colours which agreed with the observations, and also properties of the primary star which met the CCSN progenitor criteria. Both of the model primaries had $M_{\rm ini}=8$~M$_{\odot}$ and relatively massive companions (mass ratio $q=0.6$ and 0.7, respectively), although are systems with long initial orbital periods ($\sim$630 d). A look at these binary model tracks on a Hertzsprung-Russell diagram would show that they follow essentially the same evolution as the 8~M$_{\odot}$ single-star model. These binary models are thus consistent with the available evidence that low-luminosity SN~II-P progenitors, if binary, are in wide, non-interactive binaries (e.g., \citealt{ONeill2019}).

Note that we have not included circumstellar dust for any of these models, especially the RSGs, primarily since little evidence exists that such dust is important for RSGs at this low $M_{\rm ini}$ \citep[e.g.,][]{Massey2005,ONeill2019,ONeill2021}. We also cannot be certain that the presumed SAGB does not have at least some circumstellar dust as well \citep[e.g.,][]{DellAgli2017}. However, it is possible that some RSG progenitors may experience outbursts prior to explosion that may result in obscuring material that could ``hide'' the progenitor from detection, although such outbursts are predicted to occur $< 1$ yr prior to the SN (\citealt{Davies2022}; none of the archival {\sl HST\/} or {\sl Spitzer\/} data were obtained that soon before explosion). We consider the {\sl Spitzer\/} upper limits, above, and the implications for the possibility of an undetected SN 2022acko progenitor detection, neighboring the candidate within its PSF (i.e., that the candidate is unrelated to the actual progenitor). At the assumed host distance (and with negligible extinction), these limits correspond to absolute brightnesses $> -13.3$ mag in both {\sl Spitzer\/} bands. Note that the luminous, dusty progenitor of SN 2017eaw \citep[e.g.,][]{VanDyk2019,VanDyk2023} had absolute brightnesses of $-11.46$ and $-11.67$ mag in the two respective bands (not shown in Figure~\ref{fig:progenitor_sed}). If there were a dusty progenitor for SN 2022acko, it could have had $\gtrsim 1.6$ mag more circumstellar dust extinction than did the progenitor of SN 2017eaw and still evaded detection by {\sl Spitzer}; such dust would likely have been destroyed during the SN explosion.

\section{A supernova-based distance and implications for the progenitor mass}\label{sec:SCM}

The results of our analysis, above, are highly dependent on our assumption of the distance to the host galaxy. If we could somehow obtain an independent estimate of the distance, then that might bolster (or diminish) our confidence in these progenitor initial mass constraints. Fortunately, it is possible to derive such an estimate from the SN itself, in the form of the ``standardized candle method'' \citep[SCM; e.g.,][]{Hamuy2002,Nugent2006} for SNe II-P. The SCM is based on measurements of both the $I$-band apparent brightness and the SN expansion velocity from the Fe~{\sc ii} $\lambda$5169 line at day 50 (approximately the midway point in an SN~II light-curve plateau): $I_{50}$ and $v_{50}$, respectively. \citet{Polshaw2015} have calibrated the SCM technique, using several SNe~II-P which occurred in hosts having Cepheid-based distances.

\begin{figure}
	\includegraphics[width=\columnwidth]{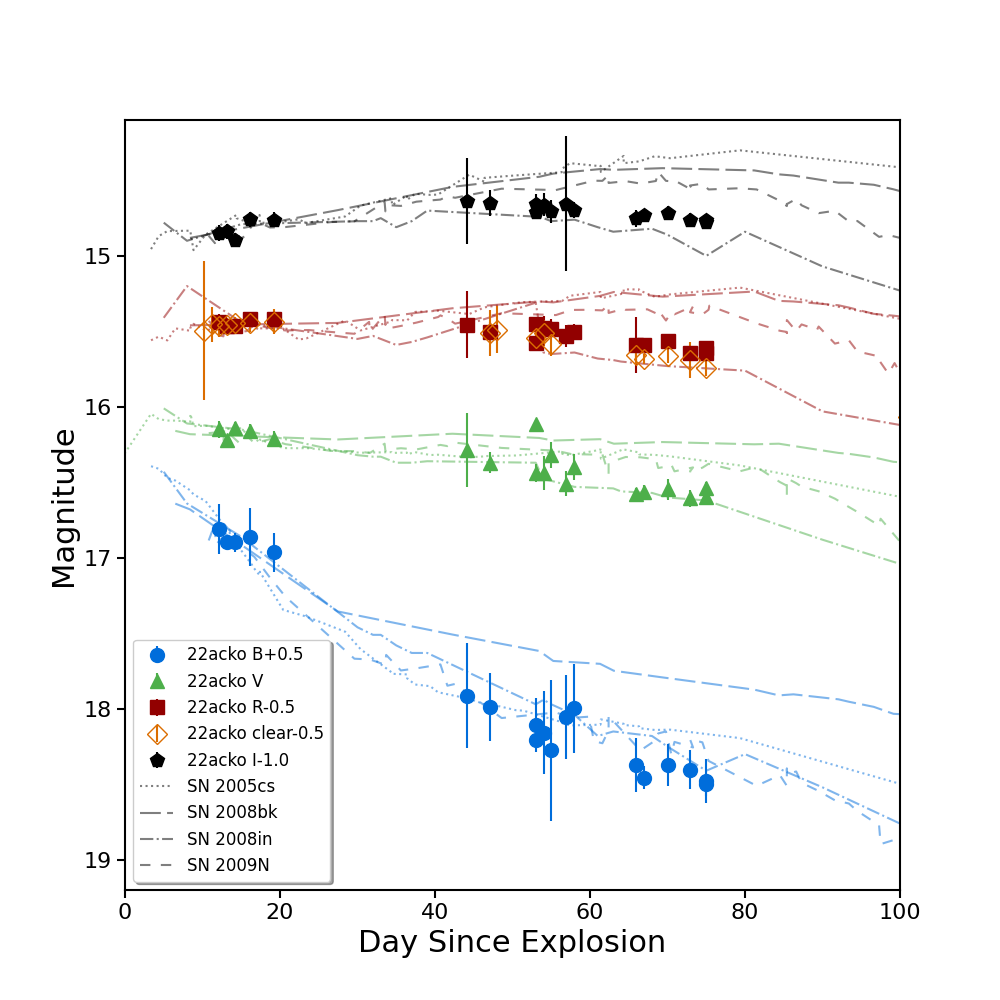}
    \caption{$BVRI$ and unfiltered (clear) light curves of SN~2022acko assembled from Lick Observatory KAIT and Nickel photometry, shown with respect to the explosion epoch JD~2,459,918.67 \citep{Bostroem2023}; see Table~\ref{tab:sn2022acko_phot}. Note that the photometry presented here has not been template-subtracted. Also shown for comparison are light curves for the low-luminosity SN~2005cs \citep{Pastorello2009}, SN~2008bk \citep{Pignata2013}, SN~2008in \citep{Roy2011}, and SN~2009N \citep{Takats2014}. SN~2022acko, at least at these early times, appears to most closely resemble SN~2008in, in terms of light-curve shape.}
    \label{fig:light_curve}
\end{figure}

In addition to the SN observations which \citet{Bostroem2023} presented, we had also conducted early-time photometric and spectroscopic monitoring of SN~2022acko at Lick Observatory (on Mt. Hamilton, CA, USA) with the 0.76~m Katzman Automatic Imaging Telescope (KAIT), the 1~m Nickel telescope, and the 3~m Shane telescope equipped with the Kast double spectrograph. Details regarding the facilities and the reduction methods for the photometric data can be found in \citet{Stahl2019} and \citet{Zheng2022}, and for the spectroscopic data in \citet{deJaeger2019}. Our coverage of the SN, unfortunately, was hindered by stretches of unusually inclement winter weather at the observatory. Nevertheless, we obtained photometry, specifically in $I$, which straddle day 50, as well as a spectrum from day 54. (Here, we have adopted the explosion epoch, JD 2,459,918.67, from \citealt{Bostroem2023}.) See Figures~\ref{fig:light_curve} and \ref{fig:spectra} for the Lick light curves and spectra, respectively, and Table~\ref{tab:sn2022acko_phot} in Appendix~\ref{sec:Lick_data} for the photometry.

For $I_{50}$ we interpolate between the surrounding values on days 47.1 and 53.1 and estimate it to be $15.65 \pm 0.07$~mag. For $v_{50}$ we measure $2,845 \pm 34$~km~s$^{-1}$ from the day 54.0 spectrum and assume that the velocity has not changed significantly from day 50. The extinction is $A_I=0.13$~mag for $R_V=3.1$. Adopting the \citet{Polshaw2015} calibration, we arrive at an SCM distance of $23.4 \pm 3.9$~Mpc ($31.84 \pm 0.37$~mag; the uncertainties in the distance estimate arise from the measurement uncertainties and those in the \citealt{Polshaw2015} calibration), which agrees with the NAM distance we assumed above to within the uncertainties. 

\begin{figure}
	\includegraphics[width=\columnwidth]{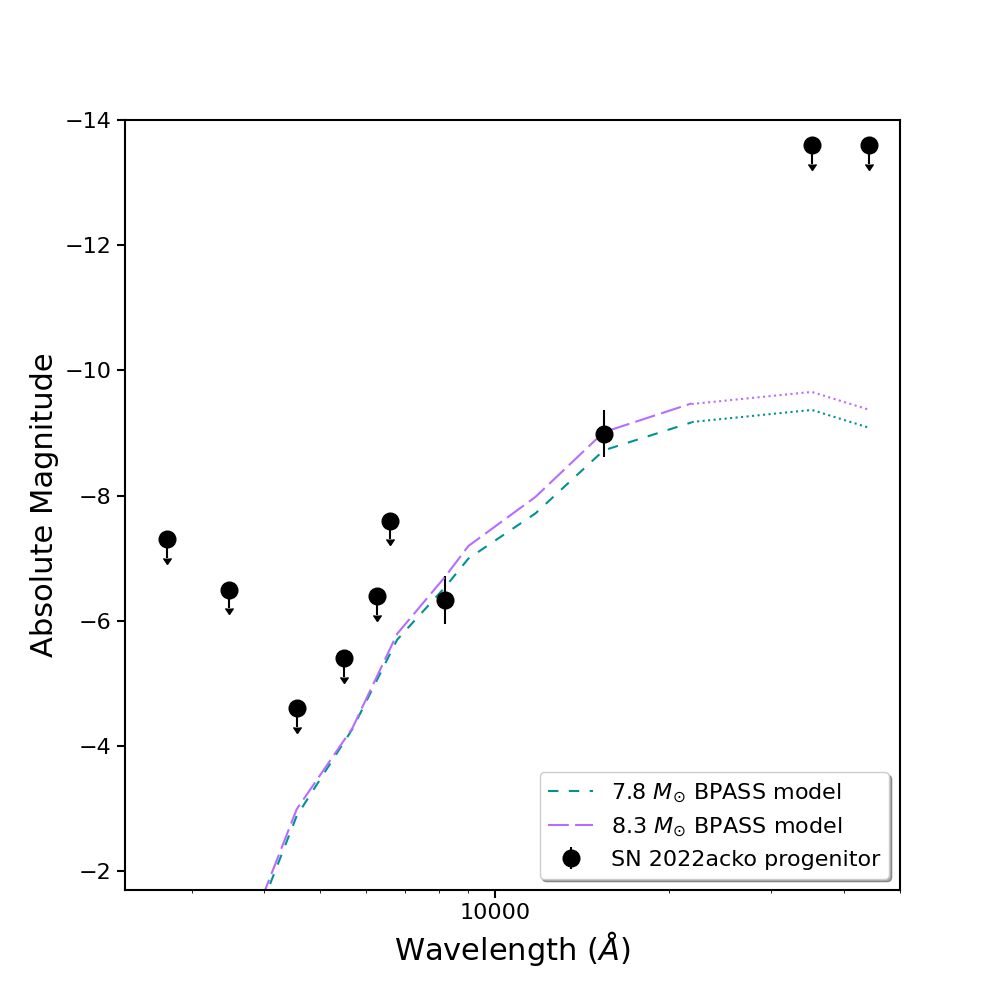}
    \caption{Same as Figure~\ref{fig:progenitor_sed}, although assuming a distance of $23.38 \pm 3.90$~Mpc, based on an SCM estimate from SN~2022acko (see Sec.~\ref{sec:Lick_data}). Also shown are BPASS single-star model endpoint SEDs at $M_{\rm ini}=7.8$ and 8.3~M$_{\odot}$, with the SEDs extended using MARCS models in a similar fashion.}
    \label{fig:progenitor_sed2}
\end{figure}

Figure~\ref{fig:progenitor_sed2} is essentially the same as Figure~\ref{fig:progenitor_sed}, except with the observed data points and limits shifted to somewhat higher luminosities, based on the SCM distance. Within the uncertainties, the data are consistent with slightly more-massive single-star models, with $M_{\rm ini} = 7.8$--8.3~M$_{\odot}$. Binary models with primaries of $M_{\rm ini} = 8$--9~M$_{\odot}$, with a range of $q$ and long initial periods of $\sim$631--10,000~d, also are consistent with the observations. Overall, these results are very similar to those in Sec.~\ref{sec:progenitor}, where we assumed the NAM distance. The moderately higher luminosities do, however, also open up other interesting binary-star possibilities: Very short-period ($<2$~d) systems consisting of $M_{\rm ini} = 4.5$--7~M$_{\odot}$ primaries, which seemingly accrete most or all of their secondary's mass, ultimately ascending a RSG branch and putatively exploding, much like a single 8~M$_{\odot}$ star.

\begin{figure}
	\includegraphics[width=\columnwidth]{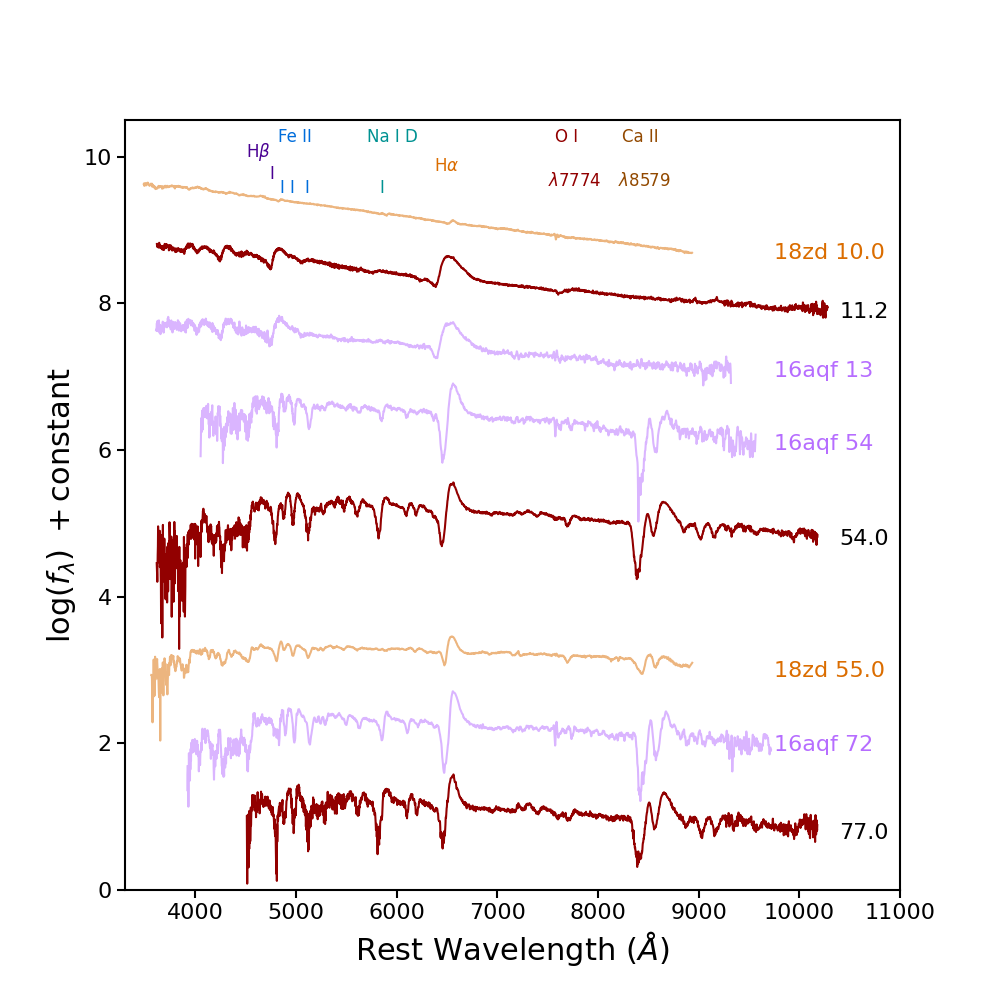}
    \caption{Spectra of SN~2022acko obtained with the Lick Observatory Kast spectrograph on 2022 December 16.33 (day 11.2), 2023 January 28.14 (day 54.0), and 2023 February 20 (day 77.0). Also shown for comparison are spectra of the low-luminosity SN~2016aqf \citep{Muller2020} and SN~2018zd \citep{Zhang2020} at similar epochs. SN~2022acko exhibits, in the last two epochs, the prominent Ca~{\sc ii} $\lambda\lambda$8498, 8542, 8662 near-infrared triplet feature typical of low-luminosity SNe~II-P. Telluric absorption has been removed through comparison with standard stars that were observed for relative-flux calibration.}
    \label{fig:spectra}
\end{figure}

\section{Discussion and conclusions}

We have, for the first time, identified an SN progenitor candidate in pre-explosion {\sl HST\/} archival data using {\sl JWST}, but only by happenstance. The star that exploded as SN~2022acko in NGC~1300 was possibly an RSG with $M_{\rm ini} \approx 7.7$~M$_{\odot}$. In our view, the fact that the F814W and F160W detections, considered together, are consistent with an RSG SED, as we would expect for an SN~II-P progenitor, is sufficiently compelling to believe that the progenitor has indeed been identified. The inferred $M_{\rm ini}$ of the candidate is slightly below the generally-recognised theoretical juncture at $\sim 8$~M$_{\odot}$ for core collapse, as opposed to WD formation in less-massive stars. However, this is also just slightly above what the models we considered predict to be an SAGB star, which presumably might end its life as an ECSN. We therefore conclude that SN~2022acko is most likely a normal SN~II-P, as the observations tend to demonstrate  (\citealt{Bostroem2023} and this paper). The slight discrepancy in our astrometric results, matching the {\sl HST\/} data to the {\sl JWST} observations and WFC3/IR to WFC/UVIS, is motivation itself for future {\sl HST\/} follow-up imaging, when the SN has significantly faded, to confirm that the candidate was indeed the progenitor --- this should include when the SN is fainter, although still detectable, to verify the astrometric association of the SN with the progenitor candidate, and when the SN has faded below the original luminosity of the candidate, to demonstrate that the candidate has vanished. Such is good practice, in general, for all progenitor identifications (e.g., \citealt{Maund2009,Maund2014,Maund2015,VanDyk2023}).

With further analysis we do not expect estimates of the interstellar reddening to the SN to evolve by much (all indications are that the total reddening is relatively low); thus, the shape of the progenitor candidate's SED should not change. However, if a more reliable distance could be measured (e.g., using Cepheids), the overall estimated luminosity of the star could be revised. We would not expect the luminosity estimate to get any lower (the star already likely sits at the RSG/SAGB boundary); however, a larger distance would only increase the star's overall luminosity and, consequently, the estimate of its $M_{\rm ini}$ as well. Our attempt to estimate an SCM distance, based on the available early-time data for the SN, may indicate that the SN is slightly farther than the assumed NAM distance, implying that a somewhat higher $M_{\rm ini}$ of $\sim$8--9~M$_{\odot}$ is also possible.

The present work launches an exciting new avenue for future potential SN progenitor identifications via {\sl JWST\/}, either through further serendipity or dedicated guest investigator or discretionary observations. The remarkable spatial resolution and sensitivity of NIRCam, relative to {\sl HST}, are attractive factors in this regard. Of concern, though, is any residual astrometric distortion in the NIRCam data, as well as possible saturation, because of the unprecedented sensitivity, if a nearby SN is observed when is still quite bright. The first effect might be mitigated against if a subarray mode were employed, rather than the full array, and the second effect could be avoided if {\sl JWST\/} observations were scheduled at later times in the SN's photometric evolution. A non-disruptive target-of-opportunity observation (with activation time greater than 14~d) might be appropriate for this reason. The short-wavelength channel would also likely be the superior choice, given its higher spatial resolution and proximity in wavelength range (0.6--2.3~$\mu$m) to that of {\sl HST}. As the {\sl JWST\/} mission matures, and more nearby host galaxies are observed in a number of bands, it may well soon be possible to identify with {\sl JWST\/} SN progenitor candidates having SEDs based on {\sl JWST\/} plus {\sl HST\/} data together or on {\sl JWST\/} data alone.

\section*{Acknowledgements}

We thank the anonymous reviewer for helpful comments that improved this paper.
This work is based in part on observations made with the NASA/ESA {\it Hubble Space Telescope} obtained from the Space Telescope Science Institute (STScI), which is operated by the Association of Universities for Research in Astronomy (AURA), Inc., under the National Aeronautics and Space Administration (NASA) contract NAS 5-26555. This work is also based in part on observations made with the NASA/ESA/CSA {\sl James Webb Space Telescope}. The data were obtained from the Mikulski Archive for Space Telescopes at STScI. This research has made use of the NASA/IPAC Infrared Science Archive, which is funded by NASA and operated by the California Institute of Technology.
This research made use of {\tt photutils}, an Astropy package for detection and photometry of astronomical sources \citep{Bradley_2022}.
This publication was made possible through the support of an LSSTC Catalyst Fellowship to K.A.B., funded through grant 62192 from the John Templeton Foundation to LSST Corporation. The opinions expressed in this publication are those of the authors and do not necessarily reflect the views of LSSTC or the John Templeton Foundation.
Time-domain research by the University of Arizona team and D.J.S. is supported by the National Science Foundation (NSF) grants AST-1821987, 1813466, 1908972, \& 2108032, and by the Heising-Simons Foundation under grant \#20201864. 
J.E.A. is supported by the international Gemini Observatory, a program of NSF's NOIRLab, which is managed by AURA under a cooperative agreement with the NSF, on behalf of the Gemini partnership of Argentina, Brazil, Canada, Chile, the Republic of Korea, and the United States of America.
A.V.F.'s supernova group at U.C.~Berkeley has received financial assistance from the Christopher R. Redlich Fund, Alan Eustace (W.Z. is a Eustace Specialist in Astronomy), Frank and Kathleen Wood (T.G.B. is a Wood Specialist in Astronomy),  numerous other individual donors, and NASA/{\sl HST} grant AR-14259 from STScI. We also acknowledge the assistance of UC Berkeley students Efrain Alvarado, Ivan Altunin, Kate Bostow, Kingsley Ehrich, Cooper Jacobus, Connor Jennings, Sophia Risin, Gabrielle Stewart, and Edgar Vidal with Lick Nickel observations.
Research by S.V. and Y.D. is supported by NSF grant AST-2008108.

\section*{Data Availability}

All of the {\sl HST\/} and {\sl JWST\/} data analysed herein are publicly available via the Mikulski Archive for Space Telescopes (MAST) portal, https://mast.stsci.edu/search/ui/\#/hst. All of the {\sl Spitzer\/} data are publicly available via the NASA/IPAC Infrared Science Archive (IRSA), https://irsa.ipac.caltech.edu/. All of the {\sl HST\/} and {\sl Spitzer\/} photometric results that we have obtained from these data are listed above. We provide the Lick photometry in Table~\ref{tab:sn2022acko_phot} of Appendix~\ref{sec:Lick_data}. The Lick spectra have been posted to the public WISeREP site \citep{Yaron2012}.


\bibliographystyle{mnras}
\bibliography{sn2022acko_progenitor}


\appendix

\section{Lick Observatory photometry of SN 2022\lowercase{acko}}\label{sec:Lick_data}

We present in Table~\ref{tab:sn2022acko_phot} the optical ground-based photometry obtained at Lick Observatory with the 0.76~m KAIT and the 1~m Nickel telescope. The magnitudes are in the Vega system. The photometry presented here has not been template-subtracted to remove the host-galaxy contribution.

\begin{table*}
\centering
\caption{Lick Observatory photometry of SN 2022acko$^a$}
\label{tab:sn2022acko_phot}
\begin{tabular}{lccccccc} 
\hline
Date & MJD$^b$ & $B$ & $V$ & $R$ & Clear$^c$ & $I$ & Source$^d$ \\
     &     & (mag) & (mag) & (mag) & (mag) & (mag) &  \\
\hline
2022 Dec 15.36 & 59928.36 &   ...       &   ...       &   ...       & 15.99(0.46) &   ...       & KAIT \\
2022 Dec 16.36 & 59929.36 &   ...       &   ...       &   ...       & 15.95(0.11) &   ...       & KAIT \\
2022 Dec 17.31 & 59930.35 & 16.31(0.17) & 16.15(0.05) & 15.94(0.05) & 15.96(0.07) & 15.85(0.05) & KAIT \\
2022 Dec 18.32 & 59931.32 & 16.40(0.04) & 16.22(0.03) & 15.96(0.03) & 15.96(0.05) & 15.84(0.05) & KAIT \\
2022 Dec 19.33 & 59932.33 & 16.40(0.06) & 16.14(0.04) & 15.96(0.04) & 15.94(0.04) & 15.89(0.05) & KAIT \\
2022 Dec 21.34 & 59934.34 & 16.36(0.19) & 16.16(0.05) & 15.92(0.04) & 15.95(0.06) & 15.76(0.05) & KAIT \\
2022 Dec 24.34 & 59937.34 & 16.46(0.13) & 16.21(0.05) & 15.91(0.05) & 15.94(0.08) & 15.76(0.05) & KAIT \\
2022 Dec 25.31 & 59938.31 &   ...       &   ...       &   ...       &   ...       &   ...       & KAIT \\
2023 Jan 18.25 & 59962.25 & 17.41(0.35) & 16.29(0.25) & 15.96(0.22) &   ...       & 15.64(0.28) & KAIT \\
2023 Jan 21.21 & 59965.21 & 17.49(0.23) & 16.37(0.07) & 16.01(0.07) & 16.01(0.15) & 15.65(0.09) & KAIT \\
2023 Jan 22.14 & 59966.14 &   ...       &   ...       &   ...       & 15.99(0.15) &   ...       & KAIT \\
2023 Jan 23.13 & 59967.13 &   ...       &   ...       &   ...       &   ...       &   ...       & KAIT \\
2023 Jan 27.14 & 59971.14 & 17.71(0.04) & 16.11(0.03) & 16.07(0.02) &   ...     & 15.71(0.02) & Nickel \\
2023 Jan 27.17 & 59971.17 & 17.61(0.18) & 16.44(0.06) & 15.95(0.05) & 16.05(0.05) & 15.66(0.07) & KAIT \\
2023 Jan 28.19 & 59972.19 & 17.66(0.27) & 16.44(0.11) & 15.97(0.07) & 16.01(0.09) & 15.66(0.08) & KAIT \\
2023 Jan 29.11 & 59973.11 & 17.78(0.46) & 16.32(0.09) & 15.98(0.06) & 16.08(0.08) & 15.71(0.08) & KAIT \\
2023 Jan 31.11 & 59975.11 & 17.55(0.28) & 16.51(0.08) & 16.03(0.07) &   ...       & 15.66(0.45) & Nickel \\
2023 Feb 01.12 & 59976.12 & 17.50(0.29) & 16.40(0.08) & 16.00(0.05) &   ...       & 15.69(0.05) & KAIT \\
2023 Feb 09.11 & 59984.11 & 17.87(0.18) & 16.58(0.04) & 16.09(0.19) & 16.16(0.06) & 15.75(0.06) & KAIT \\
2023 Feb 10.17 & 59985.17 & 17.95(0.08) & 16.56(0.04) & 16.09(0.03) & 16.18(0.03) & 15.73(0.04) & KAIT \\
2023 Feb 13.18 & 59988.18 & 17.87(0.14) & 16.55(0.07) & 16.07(0.04) & 16.16(0.04) & 15.71(0.04) & KAIT \\
2023 Feb 16.14 & 59991.14 & 17.90(0.13) & 16.61(0.06) & 16.14(0.04) & 16.19(0.12) & 15.76(0.04) & KAIT \\
2023 Feb 18.13 & 59993.13 & 17.98(0.15) & 16.54(0.04) & 16.14(0.04) & 16.24(0.06) & 15.76(0.04) & KAIT \\
2023 Feb 18.14 & 59993.14 & 18.00(0.08) & 16.59(0.03) & 16.11(0.02) &   ...       & 15.77(0.03) & Nickel \\
2023 Mar 16.14 & 60019.14 & 18.37(0.61) & 17.19(0.20) & 16.35(0.11) & 16.56(0.20) & 15.96(0.09) & KAIT \\
\hline
\end{tabular}
\begin{tablenotes}
\item[]{\it Notes}. $^a$Magnitudes are in the Vega system, and the $1\sigma$ uncertainties are in parentheses; $^b$MJD is Modified Julian Date;
\item[]$^c$Clear is unfiltered photometry, but very similar to the $R$ band for most sources \citep{Li2003}; $^d$KAIT is the 0.76~m Katzman Automatic Imaging Telescope and Nickel is the 1~m Nickel telescope.
\end{tablenotes}
\end{table*}

\bsp	
\label{lastpage}
\end{document}